\documentclass{iclp15TCtlp}
\usepackage{mathptmx}

\usepackage{theorem}
\usepackage{amsfonts,amsmath,latexsym,amssymb}
\usepackage{tikz}
\usepackage{verbatim}
\usepackage{listings}
\usepackage{color}
\usepackage{multirow}
\usepackage{fancyvrb}

\usepackage{todonotes}
\newcommand{\pattynote}[1]{}

\usepackage{mathptmx}
\usepackage[all]{xy}
\usepackage{comment}

\usepackage{tikz}
\usetikzlibrary{circuits.ee.IEC}

\usepackage{color}

\definecolor{LightGray}{gray}{0.8}

\newcommand{\Nat}{{\mathbb N}}

\newcommand{\kw}[1]{\ensuremath{\mathsf{#1}}}

\newcommand{\cat}[1]{\ensuremath{\mathbf{#1}}}

\newcommand\bcmdtab{\noindent\bgroup\tabcolsep=0pt%
  \begin{tabular}{@{}p{10pc}@{}p{20pc}@{}}}
\newcommand\ecmdtab{\end{tabular}\egroup}

\title[Structural Resolution for Logic Programming]{Structural Resolution for Logic Programming}
\author[P.~Johann, E.~Komendantskaya and V.~Komendantskiy]
{PATRICIA JOHANN\\
Department of Computer Science, Appalachian State University, USA\\
\email{johannp@appstate.edu}\\
\vspace*{-0.1in}\and
EKATERINA KOMENDANTSKAYA\\
School of Computing, University of Dundee, UK\\
\email{katya@computing.dundee.ac.uk}\\
\vspace*{-0.1in}\and
VLADIMIR KOMENDANTSKIY\\
Moixa, UK\\
\email{vladimir@moixaenergy.com}
}
\jdate{March 2015}
\pubyear{2015}
\pagerange{\pageref{firstpage}--\pageref{lastpage}}
\doi{S1471068401001193}

\newtheorem{example}{Example}[section]
\newtheorem{definition}{Definition}[section]

\newtheorem{theorem}{Theorem}[section]

\definecolor{LightGray}{gray}{0.8}

\begin{document}

\maketitle

\begin{abstract}

\vspace*{-0.05in}
We introduce a \emph{Three Tier Tree Calculus} ($T^3C$) that defines in a systematic way 
three tiers of tree structures underlying proof search in logic programming. We use $T^3C$ to define a new -- structural -- version of resolution for logic programming.

\vspace*{-0.05in}

\end{abstract}

\begin{keywords}
Structural resolution, term trees, rewriting trees, derivation trees. 

\vspace*{-0.05in}

\end{keywords}

\section{Introduction}\label{sec:intro}

As ICLP is celebrating the 200$^{th}$ anniversary of 
George Boole, 
we are reflecting on the fundamental ``laws" underlying derivations in logic programming (LP), and making an attempt to formulate some 
fundamental principles for first-order proof search, analogous in generality to Boole's ``laws of thought"  for propositional logic~\cite{Boole}.

 




Any such principles must be able to reflect two important features of
first-order proof search in LP: its recursive and non-deterministic
nature.  For this they must satisfy two criteria: to be able to (a)
model infinite structures and (b) reflect the non-determinism of proof
search, relating ``laws of infinity" with ``laws of non-determinism"
in LP.


\vspace*{-0.1in}

\begin{example}
\label{ex:nat} 
The program $P_1$ inductively defines the set of natural numbers:
\[\begin{array}{lrll}
0. & \mathtt{nat(0)} & \gets & \\ 
1. & \mathtt{nat(s(X))}  & \gets & \mathtt{nat(X)} \\
\end{array}\]

\vspace*{0.05in}\noindent
To answer the question ``{\em Does $P_1 \vdash \mathtt{nat(s(X))}$
  hold?}'', we first represent it as the LP
query $?  \gets \mathtt{nat(s(X))}$ 
and then use SLD-resolution to resolve this query with
$P_1$.  The topmost
clause selection strategy first resolves $\mathtt{nat(s(X))}$ with $P_1$'s
second clause (Clause 1), and then resolves the resulting term with
$P_1$'s first clause (Clause 0). This gives the derivation
$\mathtt{nat(s(X))} \rightarrow \mathtt{nat(X)} \rightarrow
\mathtt{true}$, which computes the solution $\{\mathtt{X \mapsto 0}\}$
in its last step. So one answer to our question is ``Yes, provided
$\mathtt{X}$ is $\mathtt{0}$.''
\end{example}

\noindent Even for this simple inductive program, there will be clause selection strategies (or clause orderings) that will result in infinite SLD-derivations. If Clause 1 is repeatedly resolved against, the infinite computation will compute the first limit ordinal.

The least and greatest Herbrand model semantics~\cite{vEmdK76,Llo87,EmdenA85} captured very well the recursive (and corecursive!) nature of LP (thus satisfying our criterion (a)).
For example, the least Herbrand model for $P_1$ is an infinite set of finite terms
$\mathtt{nat(0), \ nat(s(0)),}$ $\mathtt{nat(s(s(0))), \ldots}$.
The greatest
  complete Herbrand model   for program $P_1$ is the set containing all of the finite terms in
  the least Herbrand model for $P_1$ together with the first limit
  ordinal $\mathtt{nat(s(s(...)))}$.
	However, due to its declarative nature, the semantics does not reflect the operational non-deterministic nature of LP, and thus fails our criterion (b). 
	
	The operational semantics of LP has seen the introduction of a variety of tree structures reflecting the non-deterministic nature of proof search: \emph{proof trees}, 
	\emph{SLD-derivation trees}, and \emph{and-or-trees}, just to name a few.
However, these do not adequately 
	capture the infinite structures arising in LP proof search.
	It is well-known that SLD-derivations for any program $P$ are
sound and complete with respect to the least Herbrand model for
$P$~\cite{Llo87}, but this soundness and completeness
depends crucially on termination of SLD-derivations, and termination is not always available in LP proof search.
As a result, logical entailment is only semi-decidable in LP.

In one attempt to match the greatest complete Herbrand semantics for
potentially non-terminating programs, an operational counterpart ---
called \emph{computations at infinity} --- was introduced in~\cite{Llo87,EmdenA85}. The operational semantics of a
potentially nonterminating logic program $P$ was then taken to be the
set of all infinite ground terms computable by $P$ at infinity.
Computations at infinity  better capture the
computational behaviour of non-terminating logic programs, but infinite computations
do not result in implementations. This observation suggests one more
criterion:
(c) our operational semantics must be able to provide an observational (constructive) approach to potential infinity and non-determinism of LP proof search,
thus incorporating ``laws of observability".

 Coinductive logic programming (CoLP)~\cite{GuptaBMSM07,SimonBMG07} 
provides a method for terminating certain
infinite SLD-derivations (thus satisfying our criteria (a) and (c)). This is based on the principle of coinduction,
which is in turn based on the ability to finitely observe coinductive
hypotheses and succeed when coinductive conclusions are reached.
CoLP's search for coinductive hypotheses and conclusions uses a fairly
straightforward loop detection mechanism. It requires the programmer
to supply annotations classifying every predicate as either inductive
or coinductive. Then, for queries marked as coinductive, it observes
finite fragments of SLD-derivations, checks them for unifying
subgoals, and terminates when loops determined by such subgoals are
found.

The loop detection mechanism of CoLP has three major limitations, all arising from the fact that  
 it has relatively week support for analysis of various proof-search strategies and term structures arising in LP proof search (and thus for our criterion (b)).

(1) It does not work  well for cases of mixed induction-coinduction. For example, to coinductively define an infinite  stream of Fibonacci numbers, we would need to include inductive clauses
 defining addition on natural numbers. Coinductive goals will be mixed with inductive subgoals. Closing such computations by simple loop detection is problematic.

(2) There are programs for which computations at infinity \emph{produces} an infinite term, whereas CoLP fails to
find unifiable loops. 

\noindent Consider the following (coinductive) program $P_2$ that has the single clause

\vspace*{0.05in}\noindent
0. $\mathtt{from(X, scons(X,Y))} \gets \mathtt{from(s(X),Y)}$

\vspace*{0.05in}\noindent Given the query $? \gets \mathtt{from(0,
  X)}$, and writing $\mathtt{[\_,\_]}$ as an abbreviation for the
stream constructor $\mathtt{scons}$, we have that the infinite term
$t' = \mathtt{from(0,[0,[s(0),[s(s(0)),\ldots]]])}$ is computable at
infinity by $P_2$ and is also contained in the greatest Herbrand model
for $P_2$.
However, $P_2 \vdash \mathtt{from(0,X)}$ cannot be proven using the
unification-based loop detection technique of CoLP. Since the terms
$\;\mathtt{from(0,scons(0,X'))}$, $\mathtt{from(s(0),}$
$\mathtt{scons(s(0),X''))}$, $\mathtt{from(s(s(0)),}$
$\mathtt{scons(s(s(s(0))),X''')}, ...$ arising in the derivation for
$P_2$ and $? \gets \mathtt{from(0,X)}$ will never unify, CoLP will
never terminate. 

(3) CoLP fails to reflect the fact that some infinite computations are not productive, i.e., do not produce an infinite term at infinity.
The notion of productivity of corecursion is well studied in the semantics of other programming languages~\cite{EndrullisGHIK10,Agda,Coq}. 
For example,
no matter how long an SLD-derivation for the following program
$P_3$
 runs, it does not \emph{produce} an infinite term, and the resulting computation is thus coinductively meaningless:

\vspace*{0.05in}\noindent
0. $\mathtt{bad(X)} \; \gets \; \mathtt{bad(X)}$

\vspace*{0.05in}\noindent Somewhat misleadingly, CoLP's loop detection terminates with success for such programs,
thus failing to guarantee coinductive construction of infinite terms (failing criterion (a)).

Is our quest for a theory of LP satisfying criteria (a), (b), and (c) hopeless? We take a step back and 
recollect that the semantics of first-order logic and recursive schemes offers one classical approach 
to formulating structural properties of potentially infinite first-order terms.
Best summarised in \emph{``Fundamental Properties of Infinite Trees"}~\cite{Courcelle83}, the approach comes down to formulating some 
structural laws underlying first-order syntax.
It starts with definition of a \emph{tree language} as a (possibly infinite) set of sequences of natural numbers satisfying conditions of prefix-closedness and finite branching. 
	Given a first-order signature $\Sigma$ together with a countable set of variables $Var$, a first-order term tree is defined as a map from a tree language $L$ to the set $\Sigma \cup Var$. 
	Size of the domain of the  map determines the size of the term tree.
	The ``laws" are then given by imposing several structural properties: (i) in a given term tree, arities imposed by $\Sigma$ must be reflected by the branching in the underlying tree language; 
(ii)	variables have  arity $0$ and thus can only occur at leaves of the trees; and  (iii) the operation of substitution is given by replacing leaf variables with term trees.
A calculus for the operation can be formulated in terms of a suitable unification algorithm. 
We give formal definitions in Sections~\ref{sec:tl} and \ref{sec:t1}.

 We extend this elegant theory of infinite trees 
to give an operational semantics of LP that satisfies criteria (a), (b), and (c).
We borrow a few general principles  from this theory. 
Structural properties of trees (given by arity and variable constraints) and operations on trees (substitutions) are defined by means of ``structural laws" that
hold for finite and infinite trees. 
This gives us constructive approach to infinity (cf. criteria (a) and (c)). It remains to find the right kind of structures to reflect the non-determinism of proof search in LP.

Given a logic program $P$ and a term (tree) $t$, the first question we may ask is whether $t$ \emph{matches} any of $P$'s clauses.
First-order term matching is a restricted form of unification 
employed in (first-order) term rewriting systems (TRS)~\cite{Terese} and --- via pattern-matching --- in functional programming.
For our $P$ and $t$, we may proceed with term matching steps recursively, mimicking an SLD-derivation in which unification is restricted to term matching.
Consider the matching sequences for four different terms and the coinductive program $P_2$ from above:


\vspace*{-0.1in}

\begin{center}
\hspace*{-0.25in}
\begin{tikzpicture}[scale=0.30,baseline=(current bounding box.north),grow=down,level distance=20mm,sibling distance=50mm,font=\footnotesize]
  \node { $\mathtt{from(0,X)}$};
  \end{tikzpicture}\hspace*{0.2in}
\begin{tikzpicture}[scale=0.30,baseline=(current bounding box.north),grow=down,level distance=20mm,sibling distance=60mm,font=\footnotesize ]
  \node { $\mathtt{from(0,[0, X'])}$}
          child { node {$\mathtt{from(s(0),X')}$}};
  \end{tikzpicture}\hspace*{0.2in}
		\begin{tikzpicture}[scale=0.30,baseline=(current bounding box.north),grow=down,level distance=20mm,sibling distance=60mm,font=\footnotesize ]
  \node { $\mathtt{from(0,[0, [s(0), X'']])}$}
	child { node{ $\mathtt{from(s(0),[s(0),X''])}$ }
          child { node {$\mathtt{from(s(s(0)),X'')}$}}};
  \end{tikzpicture}\hspace*{0.2in}
	\begin{tikzpicture}[scale=0.30,baseline=(current bounding box.north),grow=down,level distance=20mm,sibling distance=60mm,font=\footnotesize ]
  \node { $\mathtt{from(0,[0, [s(0),[s(s(0)), X''']]])}$}
	child { node{ $\mathtt{from(s(0),[s(0),[s(s(0)), X''']])}$ }
          child { node {$\mathtt{from(s(s(0)),[s(s(0)), X'''])}$}
					   child { node {$\mathtt{from(s(s(s(0))), X''')}$}
					}}};
  \end{tikzpicture}
\end{center}

\noindent Let us call term matching sequences as above \emph{rewriting trees}, to highlight their relation to TRS.
The above sequences can already reveal some of the structural properties of the given logic program.
If $\Sigma_2$ is the signature of the program $P_2$, and if we denote all finite term trees that can be formed from this signature as $\mathbf{Term}(\Sigma_2)$, then 
a rewriting tree for $P_2$ can be defined as a map from a given tree language $L$ to $\mathbf{Term}(\Sigma_2)$. Since rewriting trees are built upon term trees, we may say that term trees give a first tier of tree structures, while the rewriting trees give a second tier of tree structures.
To formulate suitable laws for the second tier, we need to refine our notion of rewriting trees. 

\begin{figure}[!]
 \vspace*{-0.1in}
{\begin{center}
\begin{tikzpicture}[scale=0.30,baseline=(current bounding box.north),grow=down,level distance=18mm,sibling distance=50mm,font=\scriptsize]
  \node { $? \gets \mathtt{nat(s(X))}$} 
	child{ node {
      $\mathtt{nat(s(X))}$}
			child[sibling distance=20mm]{node {$X_1$}}
    child[sibling distance=100mm] { node {$\mathtt{nat(s(X))} \gets \mathtt{nat(X)}$}
		  child[sibling distance=100mm] { node {$\mathtt{nat(X)}$}
        child[sibling distance=20mm] {node {$X_2$}}
				 child[sibling distance=20mm] {node {$X_3$}}}}};
  \end{tikzpicture}
	$\hspace*{0.25in}\stackrel{X_2}{\rightarrow}$\hspace*{0.5in}
	\begin{tikzpicture}[scale=0.30,baseline=(current bounding box.north),grow=down,level distance=18mm,sibling distance=50mm,font=\scriptsize]
  \node { $? \gets \mathtt{nat(s(0))}$} 
	child{ node {
      $\mathtt{nat(s(0))}$}
			child[sibling distance=20mm]{node {$X_1$}}
    child[sibling distance=100mm] { node {$\mathtt{nat(s(0))} \gets \mathtt{nat(0)}$}
				  child[sibling distance=100mm] { node {$\mathtt{nat(0)}$}
        child[sibling distance=20mm] {node {$\mathtt{nat(0)} \gets$}}
				 child[sibling distance=50mm] {node {$X_3$}}}}};
  \end{tikzpicture}
 \end{center}}
 \vspace*{-0.1in}
 \caption{\footnotesize{The rewriting trees for $P_1$ and $?\gets
     \mathtt{nat(s(X))}$ and $?\gets \mathtt{nat(s(0))}$.
		The trees form a transition relative to the Tier 2 variable $X_2$ (shown by 	$\stackrel{X_2}{\rightarrow}$). The second tree is a successful proof for $?\gets \mathtt{nat(s(X)))}$.}\vspace*{-0.1in}}
\label{pic:tree2}
\end{figure}
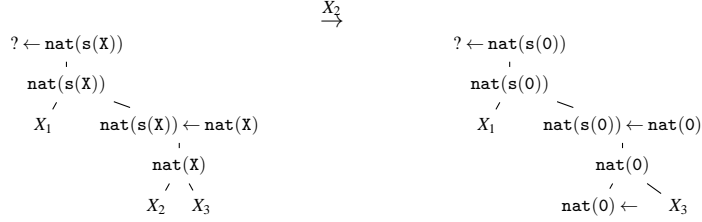

Given a program $P$ and a term $t$, 
we may additionally reflect
\emph{how many} clauses from $P$ can be unified with $t$, and how many terms those clauses contain in their bodies.
We thus introduce a new kind of ``or-nodes" to track the matching clauses.
If $P$ has $n$ clauses, $t$ may potentially have up to $n$ alternative matching sequences. 
When a clause $i$ does not match a given term tree $t$, we may use a \emph{Tier 2 variable} to denote the fact that, although $t$ does not match clause $i$ at the moment, a match may be found for some instantiation of $t$.
Thus, for the program $P_1$ above and the queries $? \gets \mathtt{nat(s(X))}$ and $? \gets \mathtt{nat(s(0))}$, we will have the two rewriting trees
of Figure~\ref{pic:tree2}. We note the alternating  or-nodes (given by clauses) and and-nodes (given by terms from clause bodies) and  Tier 2 variables.

Two kinds
of laws are imposed on structure of rewriting trees:

\begin{itemize}
\item[--] arity constraints: the arity of an and-node is
  the number of clauses in the program, and arity of and or-node is
  the number of terms in its clause body.

\item[--] variable constraints: variable leaves have arity $0$, and
  run over the objects being defined (rewriting trees). Variables are
  the leaves in which substitution can take place.
\end{itemize}

In Figure~\ref{pic:tree2}, Tier 2 variable $X_2$ is substituted by a one-node rewriting tree $\mathtt{nat(0)} \gets$.
Such substitutions 
 constitute the fundamental operation on Tier 2 trees, and give rise to a calculus for Tier 2 given in terms of so-called rewriting tree transitions. 
 Figure~\ref{pic:tree2} shows a transition from a rewriting tree for
 $? \gets \mathtt{nat(s(X))}$ to a rewriting tree for $? \gets \mathtt{nat(s(0))}$ which corresponds to the SLD-derivation outlined in Example~\ref{ex:nat}. 
Thus, a derivation is a sequence of tree transitions (given by the Tier 2 operation of substitution).
We call this method \emph{structural resolution}, or {\em S-resolution} for short. Its formal relation to TRS and type theory is given in~\cite{FK15}.
Section~\ref{sec:t2} will introduce Tier 2 formally.

We note the remarkably precise analogy between structures and operations of Tier 1 and Tier 2. 
Rewriting trees can be finite or infinite. For programs $P_1$ and $P_2$, any rewriting tree will be finite, but program $P_3$ will give rise to infinite rewriting trees. Once again, our structural analysis is fully generic for finite and infinite tree structures at Tier 2, which fits our criterion (a). Rewriting trees 
perfectly reflect the ``non-determinism laws" (criterion (b)), thanks to and-nodes and or- nodes keeping a structural account of all the search options.
Finally, our structural analysis perfectly fits criterion (c).
For productive programs like $P_1$ and $P_2$, the length of a derivation may be infinite, however, each rewriting tree will necessarily be 
 finite. This ensures observational approach to corecursion and productivity. 


We complete the picture by introducing the third tier of trees reflecting different search strategies arising from substitution into different  variables of Tier 2.
Given the set $\mathbf{Rew}(P)$ of all finite rewriting trees defined for program $P$,
a derivation tree is given by a map from a tree language $L$ to $\mathbf{Rew}(P)$. The arity of a given node in a derivation tree (itself given by a rewriting tree) is the number of
Tier 2 variables in that rewriting tree.
The construction of derivation trees is similar to the construction of SLD-derivation trees (as it accounts for all possible derivation strategies). 
The trees of Tier 3 are formally defined in Section~\ref{sec:t3}.

The resulting \emph{Three Tier Tree Calculus} ($T^3C$) developed in
this paper formalises the fundamental properties of trees arising in
LP proof search.  Apart from being theoretically pleasing, this new
theory can actually deliver very practical results.
The finiteness of rewriting trees comprising a possibly infinite
derivation gives an important observational property for defining and
semi-deciding (observational) productivity for corecursion in LP. This
puts LP on par with other languages in terms of observational
productivity and coinductive
semantics~\cite{EndrullisGHIK10,Agda,Coq}.  With a notion of
productivity in hand for LP, we can ask for results showing inductive
and coinductive soundness of derivations given by transitions among
rewriting trees. The two pictures above give, respectively, a sound
coinductive observation of a proof for $t' =
\mathtt{from(0,[0,[s(0),[s(s(0)),\ldots]]])}$ with respect to $P_2$,
and a sound inductive derivation for $\mathtt{nat(s(X))}$ with respect
to $P_1$. Our ongoing and future research based on $T^3C$ will be
further explained in Section~\ref{sec:concl}.

\vspace*{-0.25in}

\section{Background: Tree Languages}\label{sec:tl}

Our notation for trees is a variant of that in, e.g.,
\cite{Llo87,Courcelle83}. Let $\Nat^*$ denote the set of all finite
words (i.e., sequences) over the set $\Nat$ of natural numbers. The
length of a word $w\in\Nat^*$ is denoted by $|w|$. The empty word
$\epsilon$ has length $0$.  We identify the natural number $i$ and the
word $i$ of length $1$.  If $w$ is a word of length $l$, then for each
$i \in \{1,..., l\}$, $w_i$ is the $i^{th}$ element of $w$. We may
write $w = w_1...w_{l}$ to indicate that $w$ is a word of length $l$.
We use letters from the end of the alphabet, such $u,v,$ and $w$, to
denote words in $\Nat^*$ of any length, and letters from the middle of
the alphabet, such as $i,j$, and $k$, to denote words in $\Nat^*$ of
length $1$ (i.e., individual natural numbers).  The concatenation of
words $w$ and $u$ is denoted $wu$.  The word $v$ is a {\em prefix} of
$w$ if there exists a word $u$ such that $w = vu$, and a {\em proper
  prefix} of $w$ if $u \not = \epsilon$. 

\vspace*{-0.05in}

\begin{definition}\label{def:lang}
A set $L \subseteq \Nat^*$ is a \emph{(finitely branching) tree
  language} if the following conditions are satisfied:
	\vspace*{-0.05in}
\begin{itemize}
\item For all $w \in \Nat^*$ and all $i,j \in \Nat$, if $wj \in L$
  then $w \in L$ and, for all $i<j$, $wi \in L$.
\item For all $w \in L$, the set of all $i\in \Nat$ such that $wi\in
  L$ is finite.
\end{itemize}
\vspace*{-0.1in}
\end{definition}

A tree language $L$ is {\em finite} if it is
a finite subset of $\Nat^*$, and {\em infinite} otherwise.  
Examples of finite and infinite tree languages are given in
Figure~\ref{pic:tree}.
We may call a word $w \in L$ a {\em node} of $L$. If $w =
w_1w_2...w_l$, then a node $w_1w_2...w_k$ for $k < l$ is an {\em
ancestor} of $w$. The node $w$
is the \emph{parent} of $wi$,
and nodes $wi$ for $i \in \Nat$ are {\em children} of $w$.
A \emph{branch} of a tree language $L$ is a subset $L'$ of $L$ such
that, for all $w,v\in L'$, $w$ is an ancestor of $v$ or $v$ is an
ancestor of $w$. If $L$ is a tree language and $w$ is a node of $L$,
the \emph{subtree of $L$ at $w$} is $L \backslash w = \{v \mid wv \in
L\}$.  

We can now define our three-tier calculus $T^3C$.

\vspace*{-0.1in}

\section{Tier 1: Term Trees}\label{sec:t1}

In this section, we introduce Tier 1 of $T^3C$, highlighting the
structural properties of its objects (arity, branching, variables),
the operation of first-order substitution, and the relevant
calculus given by unification.

\vspace*{0.1in}

\noindent
\textbf{3.1 Tier 1 structural properties: Signature as codomain, arity, and variables}

\vspace*{0.05in}

\noindent The trees of $T^3C$'s first tier are term trees over a (first-order)
signature. A \emph{signature} $\Sigma$ is a non-empty set of
\emph{function symbols}, each with an associated 
{\em arity}. The arity of $f \in \Sigma$ is denoted $\mathit{arity}(f)$.
For example, 
$\Sigma_1 = \{\mathtt{stream}, \mathtt{scons}, \mathtt{0}\}$,
with $\mathit{arity}(\mathtt{scons}) = 2$,
$\mathit{arity}(\mathtt{stream}) = 1$, and $\mathit{arity}(\mathtt{0})
= 0$, is a signature. To define term trees over $\Sigma$, we also need
a countably infinite set $\mathit{Var}$ of {\em variables} disjoint
from $\Sigma$, each with arity $0$.  We use capital letters from the
end of the alphabet, such as $\mathtt{X}$, $\mathtt{Y}$, and
$\mathtt{Z}$, to denote variables in $\mathit{Var}$.

\begin{definition}\label{def:tt}
Let $L$ be a non-empty tree language and let $\Sigma$ be a signature.  A
\emph{term tree} over $\Sigma$ is a function $t: L \rightarrow \Sigma
\cup \mathit{Var}$ such that, for all $w\in L$, $\mathit{arity}(t(w))
= \;\,\mid \!\{i \mid wi \in L\}\!\mid$.
\end{definition}

Structural properties of tree languages extend to term trees.  For
example, a term tree $t : L \rightarrow \Sigma \cup \mathit{Var}$ has
depth $\mathit{depth}(t) = \max\{|w| \mid w\in L\}$.  The subtree of
$t$ at node $w$ is given by $t': (L \backslash
w) \rightarrow \Sigma \cup V$, where $t'(v) = t(wv)$ for each $v \in
L \backslash w$.

\begin{figure}
\begin{center}
\begin{tikzpicture}[scale=0.30,baseline=(current bounding
    box.north),grow=down,level distance=15mm,sibling
    distance=35mm,font=\scriptsize ] 
  \node { $\epsilon$}
   child {node {$0$ }
     child{ node {$00$}}
       child { node {$01$}
            }};
  \end{tikzpicture}
\hspace*{0.3in}
\begin{tikzpicture}[scale=0.30,baseline=(current bounding box.north),grow=down,level distance=15mm,sibling distance=35mm,font=\scriptsize]
  \node  {$\epsilon$}
     child { node {$0$}}
       child { node {$1$}
     child { node {$10$}}
       child { node {$\vdots$}}};
  \end{tikzpicture}
	\hspace*{0.8in}
\begin{tikzpicture}[scale=0.30,baseline=(current bounding box.north),grow=down,level distance=15mm,sibling distance=35mm,font=\scriptsize]
  \node { $\mathtt{stream}$}
   child {node {$\mathtt{scons}$ }
     child { node {$\mathtt{0}$}}
       child { node {$\mathtt{Y}$}
            }};
  \end{tikzpicture}
\hspace*{0.3in}
\begin{tikzpicture}[scale=0.30,baseline=(current bounding box.north),grow=down,level distance=15mm,sibling distance=35mm,font=\scriptsize ]
  \node  {$\mathtt{scons}$}
     child { node {$\mathtt{0}$}}
       child { node {$\mathtt{scons}$}
     child { node {$\mathtt{0}$}}
       child { node {$\vdots$}}};
  \end{tikzpicture}
 \end{center}
 \vspace*{-0.1in}
\caption{\footnotesize{The two figures on the left depict the
     finite and infinite tree languages $\{\epsilon, 0, 00, 01\}$ and
     $\{\epsilon, 0, 1, 10, 11, \ldots\}$. The two figures on the
     right depict the finite term tree $\mathtt{stream(scons(X,Y))}$
     and the infinite term tree $\mathtt{scons(0,scons(0,...))}$, both
     over $\Sigma_{1}$.}}\label{pic:tree}
\vspace*{-0.1in}
\end{figure}
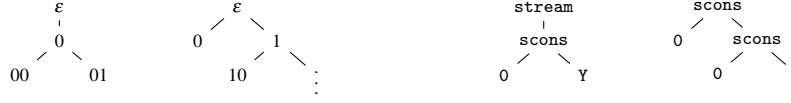

Term trees are finite or infinite according as their domains are
finite or infinite.  Term trees over $\Sigma$ may be infinite even if
$\Sigma$ is finite.  Figure~\ref{pic:tree} shows the finite and
infinite term trees $\mathtt{stream(scons(X,Y))}$ and
$\mathtt{scons(0,scons(0,...))}$ over $\Sigma_1$.  
The set of finite (infinite) term trees over a signature $\Sigma$ is
denoted $\mathbf{Term}(\Sigma)$ ($\mathbf{Term}^\infty(\Sigma)$). The
set of {\em all} (i.e., finite {\em and} infinite) term trees over
$\Sigma$ is denoted by $\mathbf{Term}^\omega(\Sigma)$. Term trees with
no occurrences of variables are \emph{ground}. We write
$\textbf{GTerm}(\Sigma)$ ($\textbf{GTerm}^\infty(\Sigma)$,
$\mathbf{GTerm}^\omega(\Sigma)$) for the set of finite (infinite, {\em
all}) ground term trees over $\Sigma$.  $\textbf{GTerm}(\Sigma)$ is
also known as the Herbrand base for $\Sigma$, and
$\mathbf{GTerm}^\omega(\Sigma)$ is known as the complete Herbrand base
for $\Sigma$, in the literature~\cite{Llo87}. Both
$\mathbf{GTerm}(\Sigma)$ and $\mathbf{GTerm}^\omega(\Sigma)$ are used
to define the Herbrand model and complete Herbrand model (declarative)
semantics of LP~\cite{Kow74,Llo87}.  Additionally,
$\mathbf{GTerm}^\omega(\Sigma)$ is used to give an operational semantics
to SLD-computations at infinity in~\cite{Llo87,EmdenA85}.

\vspace*{0.1in}

\noindent
\textbf{3.2 Tier 1 operation: First-order substitution}

\vspace*{0.05in}

\noindent A \emph{substitution} of term trees over $\Sigma$ is a total function
$\sigma: \mathit{Var} \to \mathbf{Term}(\Sigma)$.  We write
$\mathit{id}$ for the identity substitution. If $\sigma$ has finite
support --- i.e., if $|\{\mathtt{X} \in \mathit{Var}\; |\;
\sigma(\mathtt{X}) \not = \mathtt{X}\}| \in \Nat$ --- and if $\sigma$
maps the variables $\mathtt{X}_i$ to term trees $t_i$, respectively,
and is the identity on all other variables, then we may write $\sigma$
as $\{\mathtt{X}_1 \mapsto t_1,..., \mathtt{X}_n \mapsto t_n\}$. The
set of all substitutions over a signature $\Sigma$ is
$\mathbf{Subst}(\Sigma)$. Substitutions are extended from variables to
term trees homomorphically: if $t\in \mathbf{Term}(\Sigma)$ and
$\sigma \in \mathbf{Subst}(\Sigma)$, then the {\em application}
$\sigma(t)$ is defined by $(\sigma(t))(w) = t(w)$ if $t(w) \not \in
\mathit{Var}$, and $(\sigma(t))(w) = (\sigma(X))(v)$ if $w = uv$,
$t(u) = \mathtt{X}$, and $\mathtt{X} \in \mathit{Var}$.  
Composition of substitutions is denoted by juxtaposition, so
$\sigma_2\sigma_1(t)$ is $\sigma_2(\sigma_1(t))$. Since composition is
associative, we write $\sigma_3\sigma_2\sigma_1$ rather than
$(\sigma_3\sigma_2)\sigma_1$ or $\sigma_3(\sigma_2\sigma_1)$.

\vspace*{0.1in}

\noindent
\textbf{3.3 Tier 1 calculus: Unification}

\vspace*{0.05in}

\noindent A substitution $\sigma$ over $\Sigma$ is a \emph{unifier} for term
trees $t$ and $u$ over $\Sigma$ if $\sigma(t) = \sigma(u)$, and a
\emph{matcher} for $t$ against $u$ if $\sigma(t) = u$.  A substitution
$\sigma_1$ is {\em more general} than a substitution $\sigma_2$,
denoted $\sigma_1 \leq \sigma_2$, if there exists a substitution
$\sigma$ such that $\sigma \sigma_1(\mathtt{X}) =
\sigma_2(\mathtt{X})$ for every $\mathtt{X} \in \mathit{Var}$. A
substitution $\sigma$ is a {\em most general unifier} ({\em mgu}) for
$t$ and $u$ if it is a unifier for $t$ and $u$, and is more general
than any (other) such unifier. A {\em most general matcher} ({\em
  mgm}) is defined analogously.  Both mgms and mgus are unique up to
variable renaming. 

We write $t \sim_\sigma u$ if $\sigma$ is a mgu for $t$ and $u$, and
$t \prec_\sigma u$ if $\sigma$ is a mgm for $t$ against $u$. Our
notation is reasonable: unification is reflexive, symmetric, and
transitive, but matching is reflexive and transitive only. Mgms and
mgus can be computed using Robinson's seminal unification algorithm
(see, e.g.,~\cite{Llo87,Pfen06}).  Any standard unification algorithm
(possibly represented by system of sequent-like
rules~\cite{Pfen06,FK15}) can be seen as the calculus of Tier 1.
Additional details about unification and matching can be found in,
e.g.,~\cite{BS01}.


\vspace*{-0.1in}

\section{Tier 2: Rewriting Trees}\label{sec:t2}

In this section, we introduce Tier 2 of $T^3C$, highlighting the
structural properties of rewriting trees: codomains comprising term
trees and clauses, suitable notions of arity, the operation of Tier 2
substitution, and the relevant calculus given by rewriting tree
transitions.

\vspace*{0.1in}

\noindent
\textbf{4.1 Tier 2 structural properties: Terms and clauses as codomain, arity, and variables}

\vspace*{0.05in}

\noindent In LP, a {\em clause} $C$ over a signature $\Sigma$ is a pair $(A,
[B_0,...,B_n])$, where $A \in \mathbf{Term}(\Sigma)$ and $[B_0, \ldots
  B_n]$ is a list of term trees in $\mathbf{Term}(\Sigma)$.  Such a
clause $C$ is usually written as $A \gets B_0, \ldots , B_n$.  The
{\em head} $A$ of $C$ is denoted $\mathit{head}(C)$ and the {\em body}
$B_0, \ldots , B_n$ of $C$ is denoted $\mathit{body}(C)$.  In $T^3C$,
a clause over $\Sigma$ is naturally represented as a total function
(also called $C$) from a finite tree language $L$ of depth $1$ to
$\mathbf{Term}(\Sigma)$ such that $C(\epsilon) = \mathit{head}(C)$,
and if $\mathit{body}(C)$ is $B_0, \ldots , B_n$ then, for each $i \in
L$, $C(i) = B_i$.  The set of all clauses over $\Sigma$ is denoted by
$\mathbf{Clause}(\Sigma)$.  A {\em goal clause} $G$ over $\Sigma$ is a
clause $? \gets B_0, \ldots, B_n$ over $\Sigma \cup \{?\}$. Here, $?$
is a specified symbol not occurring in $\Sigma \cup \mathit{Var}$, and
$B_0, \ldots , B_n$ are term trees in $\mathbf{Term}(\Sigma)$.  The
goal clause $?  \gets \;$ is called the {\em empty goal clause} over
$\Sigma$.  We consider every goal clause over $\Sigma$ to be a clause
over $\Sigma$.  The {\em arity} of a clause $A \gets B_0 , \ldots ,
B_n$ is $n+1$.  The symbol $\mathit{head}(C)(\epsilon)$ is the
\emph{predicate} of $C$.

A \emph{logic program} over $\Sigma$ is a total function from a set
$\{0,1,\dots,n\}$ $\subseteq \Nat$ to the set of non-goal clauses over
$\Sigma$. The set of all logic programs over $\Sigma$ is denoted
$\mathbf{LP}(\Sigma)$. The {\em arity} of $P\in \mathbf{LP}(\Sigma)$
is the number $|\mathit{dom}(P)|$ of clauses in $P$.


We extend substitutions from variables to clauses and programs
homomorphically. 
The variables of a clause
$C$ can be renamed with ``fresh'' variables --- i.e., with variables
that do not appear elsewhere in the current context --- to get a new
$\alpha$-equivalent clause that can be used interchangeably with
$C$. We assume variables have been thus \emph{renamed apart} whenever
convenient. Renaming apart avoids circular (non-terminating) cases of
unification and matching in LP. Under renaming, we can
always assume that a mgm or mgu of a clause and a term is {\em
  idempotent}, i.e., that $\sigma \sigma = \sigma$.

We now define the trees of Tier 2. 
 Rewriting trees allow us to simultaneously
track all matching sequences appearing in an LP derivation, and thus
to see relationships between them.  Since rewriting trees use only
matching in their computation steps, they capture theorem proving
(i.e., computations holding for {\em all} compatible term trees). By
contrast, the Tier 3 derivation trees defined in Section~\ref{sec:t3}
use full unification, and thus capture problem solving (i.e.,
computations holding only for {\em certain} compatible term trees).

We distinguish two kinds of nodes in rewriting trees: {\em and-nodes}
capturing terms coming from clause bodies, and {\em or-nodes}
capturing the idea that every term tree can in principle match several
clause heads.  We also introduce or-node variables to signify the
possibility of unification when matching of a term tree against a
program clause fails.

\begin{definition}\label{def:CT}
Let $V_R$ be a countably infinite set of variables disjoint
from $\mathit{Var}$. If $P\in \cat{LP}(\Sigma)$, $C\in
\cat{Clause}(\Sigma)$, and $\sigma \in \mathbf{Subst}(\Sigma)$ is
idempotent, then $\kw{rew}(P,C, \sigma)$ is the function
$T:\mathit{dom}(T) \rightarrow \cat{Term}(\Sigma) \cup
\cat{Clause}(\Sigma) \cup V_R$, where $\mathit{dom}(T)$ is a non-empty
tree language, satisfying the following conditions:
\begin{enumerate}
\item $T(\epsilon) = \sigma(C)  \in
\cat{Clause}(\Sigma)$ and, for all $i \in \mathit{dom}(C)
  \setminus \{\epsilon\} $, $T(i) = \sigma(C(i))$.
\item For $w\in\mathit{dom}(T)$ with $|w|$ even and $|w| > 0$, $T(w)
  \in
\cat{Clause}(\Sigma)
\cup V_R$. Moreover,\\
-- if $T(w) \in V_R$, then $\{j\mid wj \in \mathit{dom}(T)\} =
  \emptyset$, and \\
-- if $T(w) = B \in \cat{Clause}(\Sigma)$, then there exists a
  clause $P(i)$ and an mgm $\theta$ for $P(i)$ against
  $\mathit{head}(B)$.  Moreover, for every $j \in \mathit{dom}(P(i))
  \setminus \{\epsilon\}$, $wj\in \mathit{dom}(T)$ and $T(wj) =
  \sigma(\theta(P(i)(j)))$.
\item For $w\in\mathit{dom}(T)$ with $|w|$ odd, $T(w) \in
  \cat{Term}(\Sigma)$.  Moreover, for every $i\in \mathit{dom}(P)$, we
  have\\
  -- $wi\in \mathit{dom}(T)$, and\\
  -- $T(wi) =
    \begin{cases}
 \sigma(\theta(P(i))) & \text{if }
\mathit{head}(P(i)) \prec_\theta T(w) \text{ and} \\  \text{a fresh }
 X\in V_R &\text{otherwise}
    \end{cases}$
\item No other words are in $\mathit{dom}(T)$.
\end{enumerate}
A
node $T(w)$ of $\kw{rew}(P,C,\sigma)$ is an {\em or-node} if $|w|$ is
even and an {\em and-node} if $|w|$ is odd. The node $T(\epsilon)$ is
the \emph{root} of $\kw{rew}(P,C, \sigma)$. If $P \in
\cat{LP}(\Sigma)$, then $T$ is a {\em rewriting tree for $P$} if it is
either the empty tree or $\kw{rew}(P,C,\sigma)$ for some $C \in
\cat{Clause}(\Sigma)$ and $\sigma \in \mathbf{Subst}(\Sigma)$.
\end{definition}

The arity of a node $T(w)$ in $T = \kw{rew}(P,C,\sigma)$ is
$\mathit{arity}(P)$ if $T(w) \in \mathbf{Term}(\Sigma)$,
$\mathit{arity}(C)$ if $T(w) \in \mathbf{Clause}(\Sigma)$, and $0$ if
$T(w) \in V_R$.  The role of the parameter $\sigma$ in the definition
of $\kw{rew}$ will become clear when we discuss the notion of
substitution for Tier 2.  For now, we may think of $\sigma$ as the identity
substitution.

\begin{example}\label{ex:subst1}
The rewriting trees $\kw{rew}(P_1,?\gets
     \mathtt{nat(s(X))},\mathit{id})$ and $\kw{rew}(P_1,?\gets
     \mathtt{nat(s(0))},\mathit{id})$ 
are shown in
Figure~\ref{pic:tree2}. 
\end{example}



	

A rewriting tree for a program $P$ is finite or infinite according as
its domain is finite or infinite. We write $\mathbf{Rew}(P)$ for the
set of finite rewriting trees for $P$, $\mathbf{Rew}^{\infty}(P)$ for
the set of infinite rewriting trees for $P$, and
$\mathbf{Rew}^\omega(P)$ for the set of all (finite and infinite)
rewriting trees for $P$. In~\cite{KPS12-2}, a logic program $P$ is
called (observationally) productive, if each rewriting tree
constructed for it is in $\mathbf{Rew}(P)$. Programs $P_1$ and $P_2$
are productive in this sense, whereas program $P_3$ is not.  In future
work, we will introduce methods that semi-decide observational
productivity.


\vspace*{0.1in}

\noindent
\textbf{4.2 Tier 2 operation: Substitution of rewriting trees for Tier 2 variables}

\vspace*{0.05in}

With rewriting trees as the objects of Tier 2 and a suitable notion of a
Tier 2 variable, we can replay Tier 1 substitution at Tier 2 by
defining Tier 2 substitution to be the replacement of Tier 2 variables by
rewriting trees.  However, in light of the structural dependency of
rewriting trees on term trees in Definition~\ref{def:CT}, we
must also incorporate first-order substitution into Tier 2
substitution. Exactly how this is done is reflected in the next
definition.

\begin{definition}\label{def:substs}
Let $P \in \textbf{LP}(\Sigma)$, $C \in \mathbf{Clause}(\Sigma)$,
$\sigma,\sigma' \in \mathbf{Subst}(\Sigma)$ idempotent,
and $T = \kw{rew}(P,C,\sigma)$. Then the rewriting tree $\sigma'(T)$ is defined as
follows:
\begin{itemize}
\item for every $w \in \mathit{dom}(T)$ such that $T(w)$ is an
  and-node or non-variable or-node, $(\sigma'(T))(w) = \sigma'(T(w))$.
\item for every $wi \in \mathit{dom}(T)$ such that $T(wi) \in V_R$, if
  $\theta$ is an mgm of $\mathit{head}(P(i))$ against
$\sigma'(T)(w)$, then
	$(\sigma'(T))(wiv) =
  \kw{rew}(P,\theta(P(i)),\sigma'\sigma)(v)$.
	(Note $v = \epsilon$ is
  possible.)  If no mgm of $\mathit{head}(P(i))$ against $\sigma'(T)(w)$
   exists, then $(\sigma'(T))(wi) = T(wi)$.
\end{itemize}
\end{definition}

\noindent
Both items in the above definition are important in order to make sure
that, given a rewriting tree $T$ and a first-order substitution
$\sigma$, $\sigma(T)$ satisfies Definition~\ref{def:CT}.

\vspace*{-0.1in}
\begin{example}\label{ex:subst}
Consider the first rewriting tree $T$ of Figure~\ref{pic:tree2}. Given
first-order substitution $\sigma = \{X \mapsto 0\}$, the second tree
of that Figure gives $\sigma(T)$. Note that Tier 2 variable $X_2$ is
substituted by the one-node rewriting tree $\mathtt{nat(0) \gets}$ as
a result. In addition, all occurrences of the first-order variable
$\mathtt{X}$ in $T$ are substituted by $\mathtt{0}$ in $\sigma(T)$.
\end{example}

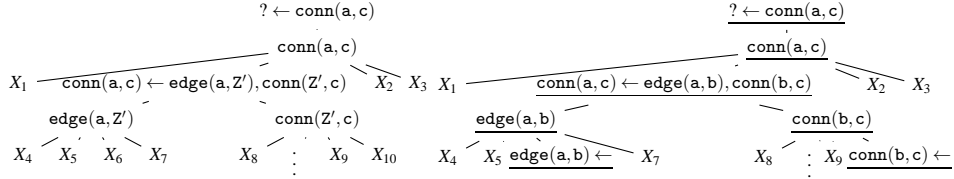
\begin{figure*}
\begin{center}
 \begin{tikzpicture}[level distance=8mm,sibling distance=50mm,scale=.6,font=\scriptsize,baseline=(current bounding box.north),grow=down]
  \node {? $\gets \mathtt{conn(a,c)}$}
		child[sibling distance=20mm]{node
                  {$\mathtt{conn(a,c)}$}
			child[sibling distance=44mm]{node {$X_1$}}
		child[sibling distance=50mm] {node
                  {$\mathtt{conn(a,c)} \gets
                    \mathtt{edge(a,Z'),conn(Z',c)}$}
	child[sibling distance=50mm]{node {$\mathtt{edge(a,Z')}$}
			child[sibling distance=10mm]{node {$X_4$}}
				child[sibling distance=10mm]{node {$X_5$}}
			   child[sibling distance=10mm]{node {$X_6$}}
	child[sibling distance=10mm]{node {$X_7$}}
			}
	child[sibling distance=50mm]{node {$\mathtt{conn(Z',c)}$}
		child[sibling distance=10mm]{node {$X_8$}}
				child[sibling distance=10mm]{node {$\vdots$}}
			   child[sibling distance=10mm]{node {$X_9$}}
	child[sibling distance=10mm]{node {$X_{10}$}}
			}}
		child[sibling distance=30mm]{node {$X_2$}}
		child[sibling distance=15mm]{node {$X_3$}}};
  \end{tikzpicture}\hspace*{-0.05in}
 \begin{tikzpicture}[level distance=8mm,sibling distance=50mm,scale=.6,font=\scriptsize,baseline=(current bounding box.north),grow=down ]
  \node {\underline{? $\gets \mathtt{conn(a,c)}$}}
		child[sibling distance=20mm]{node
                  {\underline{$\mathtt{conn(a,c)}$}}
			child[sibling distance=50mm]{node {$X_1$}}
		child[sibling distance=50mm] {node
                  {\underline{$\mathtt{conn(a,c)} \gets
                    \mathtt{edge(a,b),conn(b,c)}$}}
	child[sibling distance=70mm]{node {\underline{$\mathtt{edge(a,b)}$}}
			child[sibling distance=10mm]{node {$X_4$}}
				child[sibling distance=10mm]{node {$X_5$}}
			   child[sibling distance=20mm]{node {\underline{$\mathtt{edge(a,b)} \gets$}}}
	child[sibling distance=20mm]{node {$X_7$}}
			}
	child[sibling distance=70mm]{node {\underline{$\mathtt{conn(b,c)}$}}
		child[sibling distance=10mm]{node {$X_8$}}
				child[sibling distance=10mm]{node {$\vdots$}}
			   child[sibling distance=1mm]{node {$X_9$}}
	child[sibling distance=10mm]{node {\underline{$\mathtt{conn(b,c)} \gets$}}}
			}}
		child[sibling distance=40mm]{node {$X_2$}}
		child[sibling distance=20mm]{node {$X_3$}}};
  \end{tikzpicture}
\end{center}
 \vspace*{-0.1in}
\caption{\footnotesize{The infinite rewriting trees $T$ and $T'$ for the program
    $P_4$ of Example~\ref{ex:conn}, the clause $\mathtt{conn(a,c)}$,
    and the substitutions $\mathit{id}$ and
    $\{\mathtt{Z'}\mapsto\mathtt{b}\}$, respectively.
			$T$  offers no proof that $P_4$ logically entails $\mathtt{conn(a,c)}$,
    but the underlined steps in $T'$ comprise precisely
    such a proof. 			The figure also illustrates a transition from $T$ to $T'$ relative to variable $X_6$.}\vspace*{-0.1in}}
\label{fig:conn}
\end{figure*}

Drawing from Examples~\ref{ex:subst1} and \ref{ex:subst}, we would
ideally like to formally connect the definition of a rewriting tree
and Tier 2 substitution, and say that, given $T = \kw{rew}(P,C,id)$
and a first-order substitution $\sigma$, $\sigma(T) =
rew(P,\sigma(C),id)$. However, this does not hold in general, as was
also noticed in~\cite{KPS12-2}.  Given a clause $C = (t \gets
t_1, \ldots , t_n)$, we say a variable $\mathtt{X}$
is \emph{existential} if it occurs in some $t_i$ but not in
$t$. The presence of existential variables shows why the third parameter in
definition of $\kw{rew}$ is crucial:

\vspace*{-0.1in}
\begin{example}\label{ex:conn}
The graph connectivity program $P_4$ is given by

\vspace*{0.05in}\noindent
0. $\mathtt{conn(X,X)} \gets$\\
1. $\mathtt{conn(X,Y)} \gets \mathtt{edge(X,Z)}, \mathtt{conn(Z,Y)}$\\
2. $\mathtt{edge(a,b)} \gets$\\
3. $\mathtt{conn(b,c)} \gets$\\

\vspace*{-0.05in}
\noindent
Figure~\ref{fig:conn} shows rewriting trees $T = \kw{rew}(P_4,C,id)$ and
$T' = \kw{rew}(P_4,C,\theta)$, where $C = ? \gets \mathtt{conn(a,c)}$,
and $\theta = \{\mathtt{Z'}\mapsto\mathtt{b}\}$.
Note that $\theta(T) = T'$ but $\kw{rew}(P_4,\theta(C),id) \neq T'$.
This happens because Clause $1$ contains an existential variable $\mathtt{Z}$ in its body, and
construction of $\kw{rew}(P_4,\theta(C),id) $  fails to apply the substitution $\theta$ down the tree. 
\end{example}

Given $T = \kw{rew}(P,C,\sigma)$, for $T' = \theta(T)
= \kw{rew}(P,C, \theta\sigma)$ to hold, we must make sure that the
procedure of renaming variables apart used implicitly when computing
mgms during the rewriting tree construction is tuned in such a way
that existential variables contained in the domain of $\theta$ are
still in correspondence with the existential variables in
$\kw{rew}(P,C, \theta\sigma)$.  We achieve this by introducing a new
renaming apart convention to supplement Definition~\ref{def:CT}.
Given a program $P$ and a clause $P(i)$ with distinct existential
variables $Z_1, \ldots ,Z_n \in Var$, we impose an additional
condition on the standard \emph{renaming apart} procedure.  During the
construction of $T =\kw{rew}(P,C,\sigma)$, when an and-node $T(w)$ is
matched with $\mathit{head}(P(i))$ via $\theta$ in order to form
$T(wi) = \theta(P(i))$, $P(i)$'s existential variables $Z_1, \ldots
,Z_n$ must be renamed apart as follows:

\vspace*{0.02in}
	
-- We partition $Var$ into two disjoint sets called $V_U$ and
   $V_E$. The set $V_E$ is used to rename existential variables apart,
   while $V_U$ is used to (re)name all other variables.
	
-- Moreover, when computing an mgm $\theta$ for $T(w)$ and $(P(i))$,
	every existential variable $Z_k$ from $Z_1, \ldots ,Z_n$
	is renamed apart from variables of $T$ using the
	following indexing convention: $Z_k \mapsto E_{wi}^k$, with
	$E_{wi}^k \in V_E$.

\vspace*{0.03in}

\noindent When writing $T(wi) = \theta(P(i))$  we assume that the above
renaming convention is already accounted for by $\theta$. This ensures
that
the existential variables will be uniquely determined and synchronized
 for every two nodes $T(w)$ and $T'(w)$ in $T = \kw{rew}(P,C,\sigma)$
 and $T' = \kw{rew}(P,C, \theta\sigma)$.  Subject to this renaming
 convention, the following theorem holds.

\vspace*{-0.1in}
\begin{theorem}\label{prop:sub-props}
Let $P \in \textbf{LP}(\Sigma)$, $C \in \cat{Clause}(\Sigma)$, and
$\theta, \sigma \in \mathbf{Subst}(\Sigma)$. Then $\theta(\kw{rew}
(P, C, \sigma)) = \kw{rew} (P, C, \theta\sigma)$.
\end{theorem}\vspace*{-0.1in}
\emph{Proof.}
Let $T = \kw{rew}(P, C, \sigma)$, and let $T' = \kw{rew} (P,
C, \theta\sigma)$.  We need to prove that $\theta(T) = T'$.
The proof proceeds by induction on the length of the tree $T$ and by
cases on the types of nodes in $T$ and $\theta(T)$.

-- If $T(w)$ and $\theta(T(w))$ are non-variable or-nodes (including the
case $T(\epsilon)$), then, by Definition~\ref{def:substs},
$\theta(T)(w) = \theta(T(w)) = \theta \sigma (C^*)$, where $C^*$ is
either $C$ (i.e., it is a root node) or some $P(i) \in P$. But, by
Definition~\ref{def:CT}, $T'(w) = \theta\sigma(C^*)$.  (Here, the
synchronisation of renamed existential variables is essential, as
described.)

-- If $T(w)$ and $\theta(T(w))$ are and-nodes, then the argument is
similar.

-- If $T(wi)$ is a variable or-node, then, by
Definition~\ref{def:substs}, two cases are possible: 

\vspace*{0.02in}

(1) If no mgm for $\theta (T(w))$ and $\mathit{head}(P(i))$ exists,
then $\theta (T)(w) = \theta(T(w))$. But then no mgm for $T'(w)$ and
$\mathit{head}(P(i))$ exists either, so $T'(w) = \theta(T(w))$.

\vspace*{0.02in}

(2) If the mgm for $\theta (T(w))$ and $\mathit{head}(P(i))$ exists, then by
  Definition~\ref{def:substs}, $\theta(T)(wi) =
  \kw{rew}(P,$ $\theta'(P(i)), \theta\sigma)(\epsilon)$, where $\theta'$ is the mgm
  of $\theta(T(w))$ and $\mathit{head}(P(i))$.  The rest of the proof
  proceeds by induction on the depth of $\kw{rew}(P,\theta'(P(i)),
  \theta\sigma)$.

\vspace*{0.02in}

Base case. For the root $\theta(T)(wi) =
\kw{rew}(P, \theta'(P(i)),\theta\sigma)(\epsilon)$,
by Definition~\ref{def:CT} we have that $\kw{rew}(P,\theta'(P(i)), 
\theta\sigma)(\epsilon) =(\theta\sigma)  (\theta'(P(i)))$.
On the other hand,
Definition~\ref{def:CT} also gives that $T'(wi) =
(\theta\sigma)(\theta''(P(i)))$, where $\theta''$ is the mgm of
$T'(w)$ and $\mathit{head}(P(i))$. 
Since
$T'(w) = (\theta(T))(w)$ by the earlier argument for and-nodes,
$\theta'$ and $\theta''$ are mgus of equal term trees and
$\mathit{head}(P(i))$, so $\theta' = \theta''$.
Then
$T'(wi) = (\theta(T))(wi)$, as desired.


Inductive case.  We need only consider the situation when $T(wivj)$ is
undefined, but $(\theta(T))(wivj)$ is defined. By
Definition~\ref{def:substs}, $\theta(T)(wivj) =
\kw{rew}(P,\theta'(P(i)), \theta\sigma)(vj)$. This node can be either
an and-node, a variable or-node, or a non-variable or-node. The first
two cases are simple; we spell out the latter, more complex case only.

If $\theta(T)(wivj)$ is a non-variable or-node then, by
Definition~\ref{def:CT}, it must be $(\theta\sigma) (\theta^*
P(j))$, where $\theta^*$ is the mgm of $\theta(T)(wiv)$
and $\mathit{head}(P(j))$. 
On the other hand, Definition~\ref{def:CT} also gives that $T'(wivj) =
(\theta\sigma)(\theta^{**}(P(j)))$, where $\theta^{**}$ is the mgm of
$T'(wiv)$ and $\mathit{head}(P(j))$. 
Since $T'(wiv) = (\theta(T))(wiv)$ by the induction hypothesis,
$\theta^*$ and $\theta^{**}$ are mgms of equal term trees and
$\mathit{head}(P(j))$, so $\theta^* = \theta^{**}$.
Thus 
$T'(wivj) = (\theta(T))(wivj)$, as desired.
\vspace*{-0.242in}
\begin{flushright}
$\Box$
\end{flushright}


\vspace*{0.1in}

\noindent
\textbf{4.3 Tier 2 calculus: Rewriting tree transitions}

\vspace*{0.05in}

The operation of Tier 2 substitution is all we need to define
transitions among rewriting trees. 
Let $P \in \textbf{LP}(\Sigma)$ and $t\in \cat{Term}(\Sigma)$. If 
$\mathit{head}(P(i)) \sim_\sigma t$, then $\sigma$ is the \emph{resolvent}
of $P(i)$ and $t$. If no such $\sigma$ exists then $P(i)$ and $t$ have
{\em null resolvent}. A non-null resolvent is an \emph{internal
  resolvent} if it is an mgm of $P(i)$ against $t$, and it is
an {\em external resolvent} otherwise.

\vspace*{-0.1in}

\begin{definition}\label{def:resapp}
 Let $P \in \textbf{LP}(\Sigma)$ and $T = \kw{rew}(P,C,\sigma') \in
 \textbf{Rew}^\omega(P)$. If $X = T(wi) \in V_R$, then the rewriting
 tree $T_{X}$ is defined as follows.  If the external resolvent
 $\sigma$ for $P(i)$ and $T(w)$ is null, then $T_X$ is the empty
 tree. If $\sigma$ is non-null, then $T_X = \kw{rew}(P,
 C, \sigma \sigma')$.
\end{definition}

\noindent If $T \in \mathbf{Rew}^\omega(\Sigma)$ and $X \in V_R$, then 
the computation of $T_X$ from $T$ is denoted $\kw{Trans}(P,T,X) = T_X$. If the
other parameters are clear we simply write $T \rightarrow T_X$. The
operation $T \rightarrow T_X$ is a \emph{tree transition} for $P$ and
$C$. A {\em tree transition} for $P \in \mathbf{LP}(\Sigma)$ is a tree
transition for $P$ and some $C \in \mathbf{Clause}(\Sigma)$. A (finite
or infinite) sequence $T = \kw{rew}(P,C,\mathit{id}) \rightarrow T_1
\rightarrow T_2 \rightarrow \ldots$ of tree transitions for $P$ is a
\emph{derivation} for $P$ and $C$.
Each rewriting tree $T_i$ in the derivation is given by $\kw{rew}(P,
\, C, \,\sigma_i\ldots \sigma_2 \sigma_1)$, where $\sigma_1, \sigma_2
\ldots$ is the sequence of external resolvents associated with the
derivation.  When we want to contrast the above derivations with
SLD-derivations, we call them S-derivations, or
 derivations by \emph{structural resolution}. 

\begin{example}\label{ex:bl}
Tree transitions for $P_1$ and $P_4$ are shown in
Figures~\ref{pic:tree2} and~\ref{fig:conn}, respectively. 
\end{example}


It is our current work to prove that S-derivations are sound and
complete relative to declarative semantics of LP; see also~\cite{FK15}
for a comparative study of the operational properties of S-derivations
and SLD-derivations.

\vspace*{-0.2in}

\section{Tier 3: Derivation Trees}\label{sec:t3}

While the rewriting trees of Tier 2 capture transitions between Tier 1
term trees that depend on matching, the derivation trees of Tier 3
capture transitions between Tier 2 rewriting trees that depend on
unification. Derivation trees thus allow us to simultaneously track
all unification sequences appearing in an LP derivation. 
The {\em arity}
of a rewriting tree $T$, denoted $\mathit{arity}(T)$, is the
cardinality of the set $\mathit{indices}(T)$ of indices of variables
from $V_R$ in $T$.
There is always a bijection
$\mathit{pos}$ from $\mathit{indices}(T)$ to the (possibly infinite)
set $\mathit{arity}(T)$.


\begin{definition}\label{df:CD}
If $P\in \cat{LP}(\Sigma)$ and $C\in \cat{Clause}(\Sigma)$, the
\emph{derivation tree} $\kw{der}(P,C)$ is the function $D:
\mathit{dom}(D) \rightarrow \mathbf{Rew}^\omega(P)$ such that
$D(\epsilon) = \kw{rew}(P,C,\mathit{id})$, and if
$w\in\mathit{dom}(D)$, $i \in \mathit{arity}(D(w))$, and $i = pos(k)$,
then $wi\in\mathit{dom}(D)$ and $D(wi)$ is $\kw{Trans}(P,D(w),X_k)$.
\end{definition}

For $P \in \mathbf{LP}(\Sigma)$ and $C \in \mathbf{Clause}(\Sigma)$,
the derivation tree $\kw{der}(P,C)$ is unique up to renaming. If $P
\in \mathbf{LP}(\Sigma)$, then $D$ is a {\em derivation tree for $P$}
if it is $\kw{der}(P,C)$ for some $C \in \mathbf{Clause}(\Sigma)$. A
derivation tree is finite or infinite according as its domain is
finite or infinite. 
Inductive programs like $P_1$ and coinductive programs like $P_2$ will
have infinite derivation trees, so construction of the full derivation
trees for such programs is infeasible. Nevertheless, finite initial
fragments of derivation trees may be used to make coinductive
observations about various routes for proof search. We are currently
exploring this research direction.

\vspace*{-0.3in}

\section{Conclusions and Future Work}\label{sec:concl}

This paper gives the first fully formal exposition of the Three Tier
Tree Calculus $T^3C$ for S-resolution, relating ``laws of infinity",
``laws of non-determinism", and ``laws of observability" of
proof search in LP in a uniform, conceptual way. Implementation of derivations by S-resolution is available~\cite{sres}.

 The structural approach to LP put forth in this paper relies on the
syntactic structure of programs rather than on their (operational,
declarative, or other) semantics. In essence, it presents an LP
analogue of the kinds of reasoning that types and pattern matching
support in interactive theorem proving (ITP)~\cite{Agda,Coq}.  Further
study of this analogy is an interesting direction for future research.

Our next steps will be to formulate a theory of universal and
observational productivity of (co)recursion in LP, and to supply
$T^3C$ with semi-decidable algorithms for ensuring program
productivity (akin to guardedness checks in ITP). Formally proving
that S-resolution is both inductively and coinductively sound is
another of our current goals.

Since LP and similar automated proof search methods underlie type
inference in ITP and other programming languages, S-resolution also
has the potential to impact the design and implementation of typeful
programming languages. This is another research direction we are
currently pursuing.

\pagebreak

\bibliographystyle{acmtrans}
\bibliography{katya2}

\label{lastpage}
\end{document}